\documentclass[A4,12pt]{article}
\usepackage{times}

\usepackage[utf8]{inputenc}
\usepackage{hyperref}
\usepackage{graphicx}
\usepackage{color}
\usepackage{amssymb}
\usepackage{amsmath}
\usepackage[labelfont=bf]{caption}

\title{\textbf{CuAu, a hexagonal two-dimensional metal}} 
\author{Georg Zagler$^{1,*}$, Michele Reticcioli$^{1,2}$, Clemens Mangler$^1$, \\Daniel Scheinecker$^1$, Cesare Franchini$^{1,2,3}$, and Jani Kotakoski$^{1,*}$\\
$^1$Faculty of Physics, University of Vienna, Boltzmanngasse 5,\\ 1090 Vienna, Austria\\
$^2$Center for Computational Materials Science,\\ University of Vienna, Sensengasse 8, 1090 Vienna, Austria\\
$^3$Dipartimento di Fisica e Astronomia,\\Universit\`{a} di Bologna, 40127 Bologna, Italy\\
$^*$Email: georg.zagler@univie.ac.at (GZ),\\jani.kotakoski@univie.ac.at (JK)}
\date{}

\begin{document}
\baselineskip24pt

\maketitle 
 \begin{abstract}
\normalsize \bf \baselineskip22pt
Growth of two-dimensional metals has eluded materials scientists since the discovery of the atomically thin graphene and other covalently bound 2D materials. Here, we report a two-atom-thick hexagonal copper-gold alloy, grown through thermal evaporation on freestanding graphene and hexagonal boron nitride. The structures are imaged at atomic resolution with scanning transmission electron microscopy and further characterized with spectroscopic techniques. Electron irradiation in the microscope provides sufficient energy for a phase transformation of the 2D structure---atoms are released from their lattice sites with the gold atoms eventually forming face-centered cubic nanoclusters on top of 2D regions during observation. The presence of copper in the alloy enhances sticking of gold to the substrate, which has clear implications for creating atomically thin electrodes for applications utilizing 2D materials. Its practically infinite surface-to-bulk ratio also makes the 2D CuAu particularly interesting for catalysis applications.

\end{abstract}

\newpage

\section{Introduction}

The discovery of graphene in 2004~\cite{Novoselov2004} was quickly followed by an effort to expand the new class of two-dimensional materials, with several notable successes. For example, monolayers of elemental 2D materials~\cite{Vogt2012,Davila2014,Liu2014,Zhu2015,Mannix2015,Feng2016,Wu2017,Sun2017,Reis2017,Yuhara2019}, transition-metal mono- and dichalcogenides (reviewed recently in Refs.~\cite{Manzeli2017,Zhou2018}), as well as a number of other materials (for an overview, see Ref.~\cite{Geng2018}) have been reported. The interest in 2D materials can be traced back to the extraordinary properties of graphene~\cite{Novoselov2004,Novoselov2007,Novoselov2007a,Lee2008}, the changes in materials' properties when the dimensionality is reduced~\cite{Gibertini2019}, and the possibility for creating heterostructures with atomically thin materials without the need of epitaxy due to the weak van der Waals interaction between most 2D materials~\cite{Novoselov2016}. So far, the discovered 2D materials share covalent intralayer bonding, although some of the structures have a metallic character. While 2D structures consisting of metallic bonding have been sculpted within other structures by removing the non-metallic atoms~\cite{Wang2019}, self-assembly of Fe over small holes in graphene~\cite{Zhao2014} or thinning a bulk structure~\cite{Zhao2018}, typically through electron irradiation, no 2D structures with metallic bonding have until now been grown on weakly interacting substrates. In contrast, many of the aforementioned elemental 2D materials were synthesized by choosing a specific substrate providing a strong interaction that fosters the growth of the material~\cite{Zhu2015, Reis2017,Yuhara2019}.

Metallic bonds are nearly isotropic and thus lack the spatial preferences typical for covalent bonds. This leads to the close packed crystal structures typical for metals. Very small gold clusters provide a curious counter-example: when they contain less than 14 atoms, they tend to have flat structures. With increasing number of atoms, however, gold also turns three-dimensional~\cite{Xing2006}. This preference, together with weak bonding with other elements, means that covering surfaces with thin layers of gold presents a challenge for example when creating electric contacts on surfaces through lithography. This is typically solved using an adhesive layer, for example of chromium, between gold and the substrate.

Gold forms alloys with a number of other metals, including copper. The CuAu system is well known in bulk form and displays a variety of structures depending on stoichiometry and temperature \cite{Johansson1936}. Its stoichiometric phase forms a tetragonal (elongated face centered cubic) $L1_0$ phase. Cu as well as Au and its bimetallic alloys have recently garnered increasing interest for applications, e.g. in catalysis and plasmonics. While Cu has been recognized as a viable catalyst for some time \cite{Fujitani1994,Li2009}, for CuAu and Au this has not been the case. For structures that are reduced in some dimensions down to the nanometer regime, Au displays increased catalytic activity---most notably for CO oxidation at sub-ambient temperatures \cite{Valden1998,Haruta1997}. Recent efforts to reduce the dimensionality of Au even led to sub-nanometer thick Au nano-sheets, grown with a confining agent to non-uniform crystallinity, exhibiting pronounced catalytic performance \cite{Ye2019}. For bimetallic CuAu alloy nanoparticles only few studies on its catalytic reactivity have been reported so far. They have however indicated that it could be used in fine-tuning reactivities observed in Au nanoparticles \cite{He2014,Bracey2009}. Generally, two-dimensional materials with their high surface-to-bulk ratio represent a promising route to reduce the cost of catalysts. The size and shape of Au nanoparticles can also impact the photonic response, used for example in photoacoustic imaging \cite{Wang2012}, which is tunable with a bias voltage for particle thicknesses under some nm \cite{Maniyara2019}. The plasmonic response of low Cu-content CuAu nanocrystals was also already used in photothermal therapy \cite{He2014}, and tunable plasmon responses are predicted also for other atomically thin noble metals \cite{GarciaDeAbajo2015}.

Here we report for the first time the growth of a two-dimensional structure with metallic bonds on weakly interacting substrates. The growth of the stoichiometric CuAu alloy is carried out through thermal evaporation of gold at 1080$^\circ$C directly onto free-standing graphene and hBN membranes that contain copper from the growth of the substrate material itself. In situ cleaning of the samples and their atomic-resolution after-growth imaging were carried out with the Nion UltraSTEM 100 microscope in Vienna at 60~kV and 80~kV with a high-power laser and the high-angle annular dark field (HAADF) detector, respectively. Additional energy-dispersive X-ray spectroscopy (EDX) was carried out to confirm the elemental composition of the material. The discovered structure consists of a hexagonal two-atom-thick lattice, matching the atomic arrangement in the (111) plane of face centered cubic (fcc) lattice, where all gold atoms are within one layer and all copper atoms in another layer. Under continuous electron irradiation in the microscope, gold atoms are released and migrate on top of the structure, initiating the formation of fcc gold nanoclusters, demonstrating the weak coupling of the gold atoms with the substrate and their preference for a close-packed three-dimensional structure.
 
\section*{Results and Discussion}

\subsection*{Experimental}
Commercially available graphene and hBN monolayers, grown by chemical vapour deposition on copper substrates, were deposited on perforated SiN transmission electron microscopy grids. Such prepared freestanding 2D materials, with an abundance of Cu left over from the sample growth, served as substrates for the subsequent experiments. After preparation, the substrates were transferred into a vacuum system setup (base pressure $10^{-9}$~mbar) in Vienna \cite{Inani2019} connecting through a vacuum transfer system several independent experimental setups, including a customized~\cite{Hotz2016} Nion UltraSTEM 100 \cite{Krivanek2008} aberration-corrected scanning transmission electron microscopy (STEM) instrument and a Knudsen cell thermal evaporator. Graphene substrates were first imaged and subsequently cleaned in the microscope column with a high-power laser \cite{Tripathi2017} to obtain large areas of contamination-free monolayer graphene substrate. Spectroscopic pre-characterisation was carried out in situ with electron energy-loss spectroscopy (EELS) confirming abundant copper (see Supplementary Information Figure S1). Au was deposited onto the substrates in high vacuum at 1080$^\circ$C. The resulting structures were imaged at atomic resolution via STEM. On the contrary, hBN substrates were not cleaned prior to deposition, resulting in smaller ($\sim 5$~nm) metallic islands after Au deposition due to hydrocarbon contamination on the substrate, reducing the mobility of Au and acting as nucleation site \cite{Zagler2019}.

\subsection*{Microscopy and Spectroscopy}
\begin{figure*}[h]
	\includegraphics[width=\textwidth]{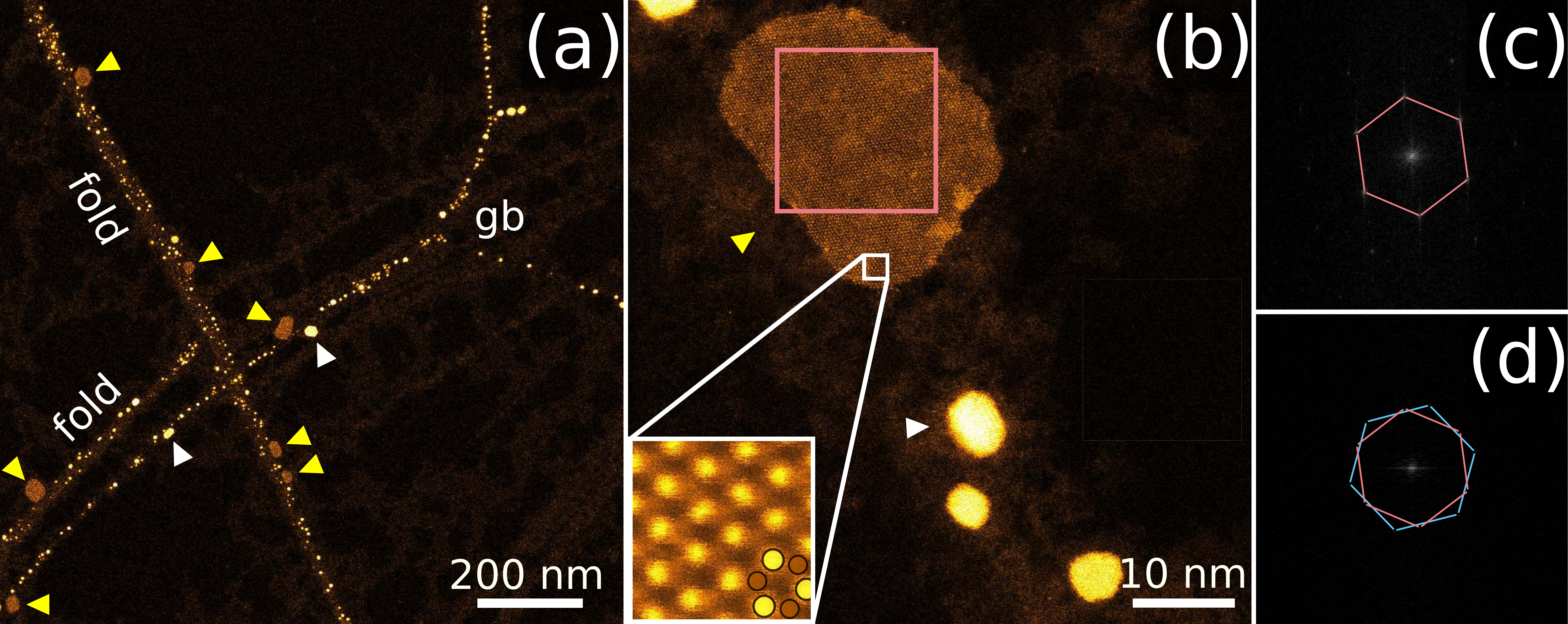}
	\caption{STEM HAADF images of CuAu on a graphene substrate.
		a) Overview of graphene substrate with nanostructures of different sizes and shapes that appear on or close to grain boundaries (gb) or folded-over graphene. The larger structures (marked by yellow triangles) display characteristic, uniform low intensities, indicating a flat structure with a comparably low thickness. Other structures (marked by white arrows) display size- and shape-dependent intensities indicating their three-dimensional nature.
		b) Higher magnification overview image shows that the uniform structures are crystalline. The inset (with overlay) reveals that they consist of two different kinds of atoms arranged into a hexagonal lattice.
		c),d) Fourier transforms taken from within the red (flat nanostructure) and blue (graphene) squares, respectively. The differing orientations of the two structures are clearly visible through the overlaid hexagons.
		All images were recorded at 60~kV.}
	\label{fig:Overview} 
\end{figure*}

\begin{figure*}[h]
	\includegraphics[width=\textwidth]{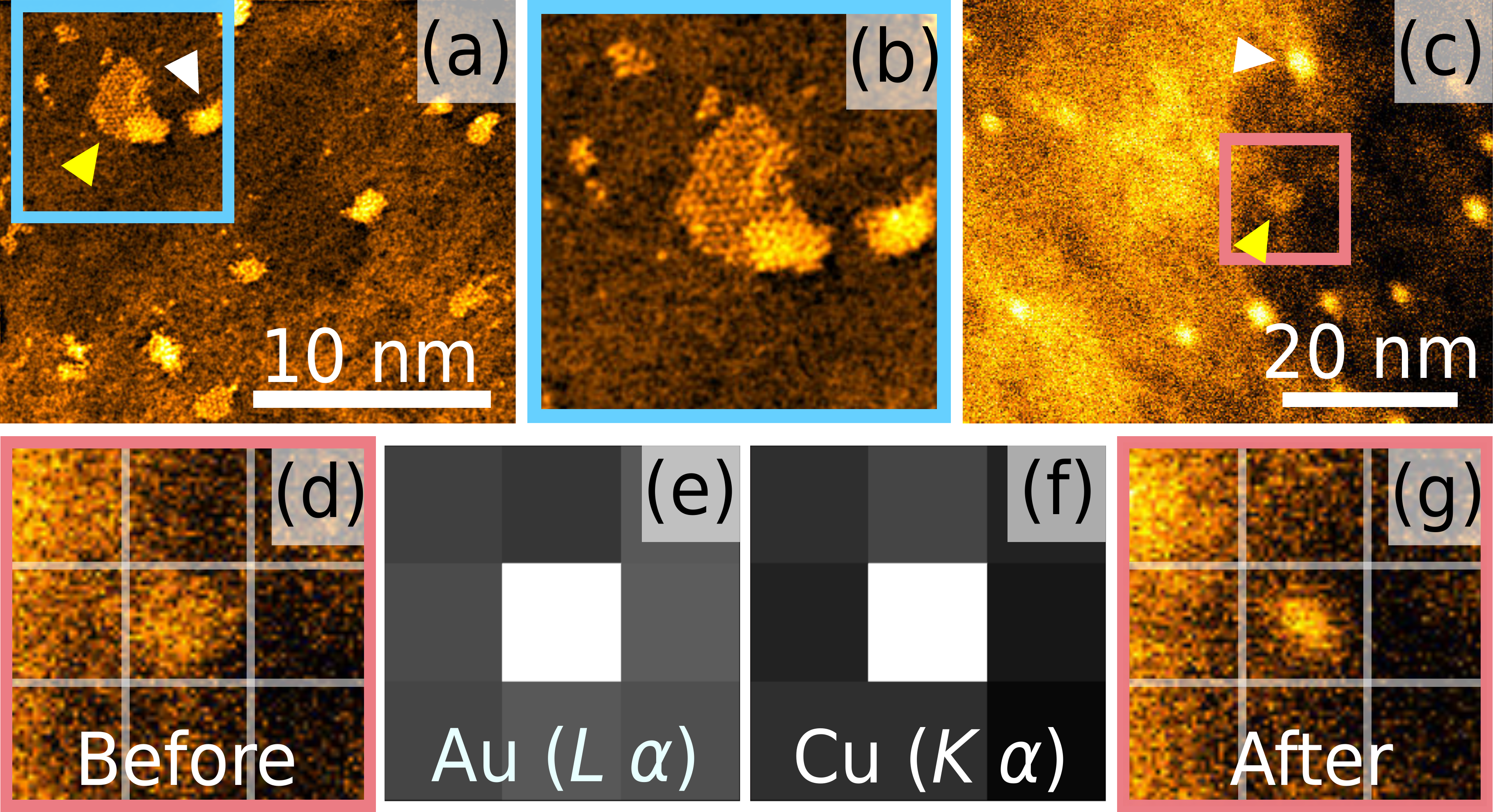}
	\caption{EDX analysis on hBN. a) STEM pre-characterization of the sample revealing both flat structures (yellow arrow) and three-dimensional clusters (white arrow), carried out at 60~kV. A magnification of the marked area is shown in (b), revealing the hexagonal lattice. c) A similar sample area imaged in STEM mode in the other microscope recorded at 80~kV. A $3\times 3$ pixel spectral map (64~s accumulation per pixel) was recorded in the marked area, shown again in (d). The corresponding integrated intensities of Au $L \alpha$ and Cu $K \alpha$ peaks are shown in (e,f), respectively. g) The same area after recording of the spectral map, revealing now a thicker structure.}
	\label{fig:EDX} 
\end{figure*}

A typical STEM HAADF image recorded at 60~kV of the metallic islands on graphene can be seen in Figure \ref{fig:Overview}a. The HAADF intensity is proportional to the atomic number of the scatterer \cite{Krivanek2010} and the thickness of the imaged location of the sample. Large areas of the graphene substrate appear dark with a uniform contrast, due to the lack of hydrocarbon contamination or metal clusters. On more active sites, such as grain boundaries (gb) or folds, contamination and brighter nanostructures can be seen. Smaller clusters (white triangle) display increasing intensity with size, indicating spherical structures (nanoclusters) for which the thickness increases with the size of the particle. Other structures similarly consisting of metal atoms are much larger in size and display uniform intensity (yellow triangle). In Figure \ref{fig:Overview}b such a larger structure is shown at higher magnification, which reveals an ordered flat arrangement of atoms with overall uniform intensity and only some thicker regions at the edges. The Fourier transforms (FT) of the structure (red box) and the underlying graphene substrate (blue box) are compared in Figure \ref{fig:Overview}c,d. The FT of the grown two-dimensional structure (Figure \ref{fig:Overview}c) reveals a hexagonal symmetry common for two-dimensional materials and in contrast to the previously reported AuC structure \cite{Westenfelder2015}. Comparing it to the graphene substrate (Figure \ref{fig:Overview}d) shows a different lattice constant (ca. 0.268~nm as compared to 0.246~nm) and orientation. This suggests that it is not epitaxial with graphene (see also the video in the Supplementary Information), typical for a weakly bound material \cite{Koma1992}. Analysis of ten images that allowed comparing the orientations of the 2D CuAu islands and the substrate revealed that there is a preferential mismatch at ca. $33^{\circ}$, although also other values were found ranging from ca. $25^\circ$ to nearly $38^\circ$ (see also the supplementary movie of the rotation of a 2D island on graphene). Increasing the magnification further (inset of Figure \ref{fig:Overview}b) reveals that the structure consist of two distinct atomic sites (columns), reflected in the different scattering intensity, arranged in a hexagonal lattice. The intensity of the brighter atom is comparable to that of individual gold atoms found outside the lattice (see also Figure \ref{fig:EDX}a, b and Figures S\ref{fig:HAADF_comparison} and S\ref{fig:Crumpling_long}). This establishes that the islands are two-dimensional with the brighter atomic column containing just one gold atom. The intensity of the other column is similar to that of individual copper atoms.

Unfortunately, the small size and electron-irradiation sensitivity of the structures as well as the overlap of gold $O_{2,3}$ and copper $M_{2,3}$ EELS peaks prevented direct in situ characterization after the growth (see Supplementary Information Figure S1). Instead, we carried out further characterization through EDX spectroscopy with an image-side corrected FEI Titan 80-300 transmission electron microscope in the STEM mode. For this, we first characterized a hBN substrate with a number of relatively small grown islands via STEM (Figure \ref{fig:EDX}a-b), and subsequently transferred the sample through air to the other microscope. The atomic resolution STEM images revealed both types of structures (larger flat ones marked with the yellow triangle as well as the smaller three-dimensional ones, white triangle). A similar sample area was then found in the other microscope, and several point spectra and a spectral map were recorded at 80~kV. The results of the $3\times 3$ pixel spectral map for the cluster marked in Figure \ref{fig:EDX}c are shown in Figure \ref{fig:EDX}d-g. The only significant peaks in the EDX spectra recorded on top of the cluster (in addition to carbon) were from Au and Cu, for which the peak intensities are shown in the maps in Figure \ref{fig:EDX}e and f, respectively (white marks the pixel with the highest intensity and other pixels are shown in gray scale reflecting the corresponding measured intensity). Clearly, the structures with uniform contrast contain only copper and gold, and those elements are limited exactly to the locations of these structures (the three-dimensional clusters contain mostly gold, but also pure copper and gold-copper clusters are found on the samples). The image recorded after the spectrum map (Figure~\ref{fig:EDX}g) reveals that the electron-beam exposure has transformed the originally flat structure into a three-dimensional nanocluster. This is a common feature for all observed structures, and will be discussed further below. Carbon is a ubiquitous contaminant, and was also found in our EDX measurements with a nm-sized probe. However, as there was no excess carbon signal at the locations of the observed structures, we assume that the islands consist only of Cu and Au, which presents the simplest model structure.

\subsection*{Structure} Atomic resolution images revealed a metallic two-component hexagonal structure that according to spectroscopic analysis consists of copper and gold atoms. To confirm that such a structure is indeed possible, we turn to density functional theory (DFT). 
DFT simulations were performed using the Vienna Ab initio Software Package (VASP) \cite{Kresse1996,Kresse1996a}, adopting the strongly constrained and appropriately normed (SCAN) functional \cite{Sun2015,Sun2016} with an energy cut-off of 500 eV. Structural optimization was carried out for the unit cell containing two atoms.
The substrate was excluded from the simulations in order to investigate the stability of the 2D CuAu structure as an independent material. This choice is further motivated by the experimental evidence of negligible interaction between the substrate and the CuAu structure.

\begin{figure*}[h]
	\includegraphics[width=\textwidth]{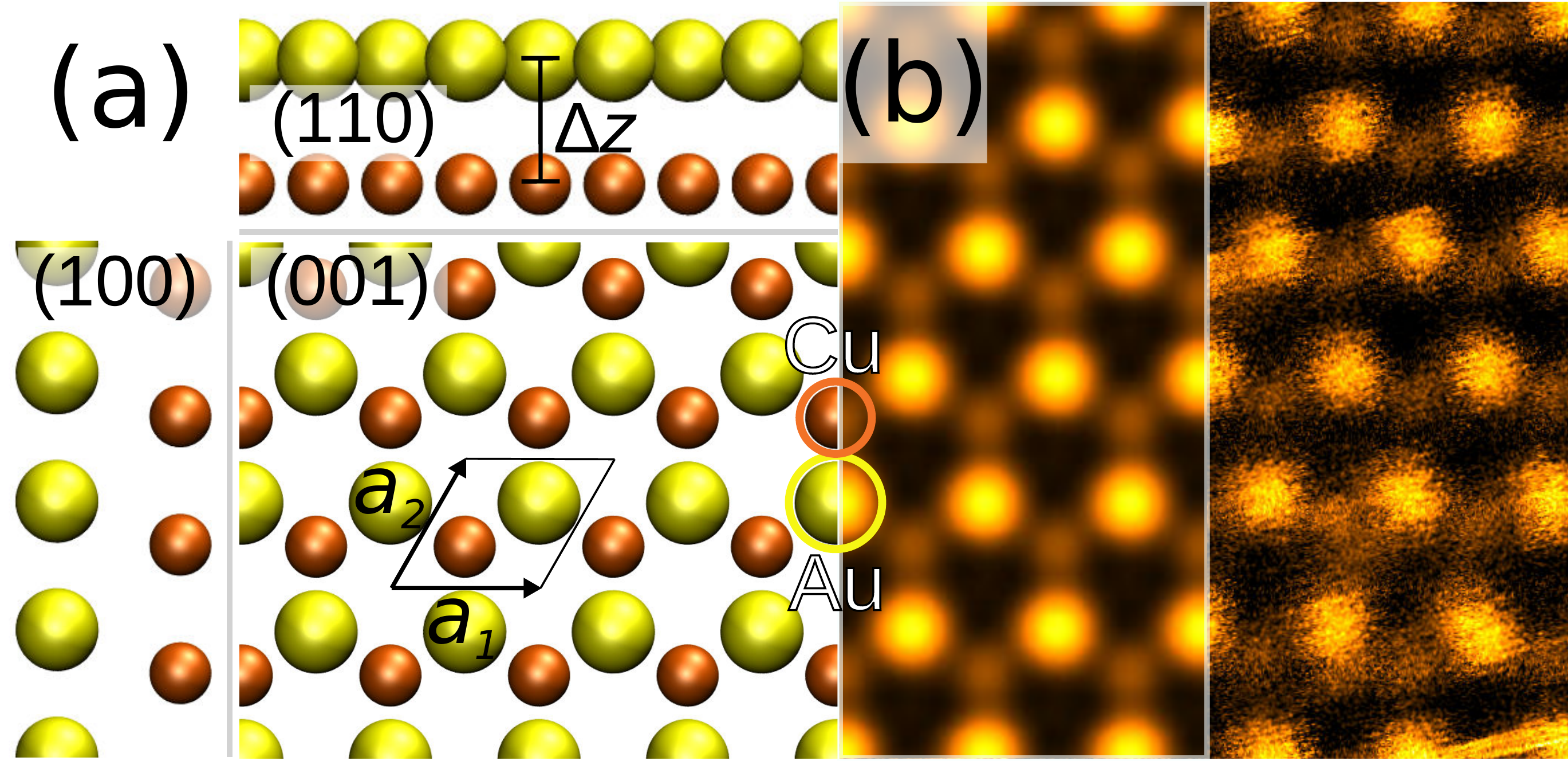}
	\caption{Atomic structure of hexagonal 2D CuAu. a) Structural model in (001), (100) and (110) crystallographic orientations. Lattice vectors $|\mathbf{a}_1| = |\mathbf{a}_2| = 0.268$~nm and $\Delta z = 0.224$~nm. b) Image simulation corresponding to the structural model (left) as compared to an experimental STEM HAADF image recorded at 60~kV (right).
	}
	\label{fig:Structure} 
\end{figure*}

The atomic structure, as relaxed by the simulations, is shown in Figure~\ref{fig:Structure}a. The in-plane lattice constant is found to be 0.268~nm, in excellent agreement with the experimental estimate of 0.268(2)~nm. The calculated inter-layer distance between the copper and gold planes is 0.224~nm. Figure~\ref{fig:Structure}b compares an image simulation (left, simulated using PyQSTEM~\cite{Koch2002,Madsen2017} with parameters corresponding to the experiments) of the thickness and atomic species dependent HAADF signal to an experimental image (right) showing again excellent agreement. This structure is reminiscent of the bulk tetragonal $L1_0$ phase cut along the (201) lattice plane and including adjacent layers of gold and copper (see Supplementary Information Figure S4), with the important difference that in order to obtain the perfect hexagonal arrangement of the atoms, a $c/a = 1/\sqrt{2}$ is required, instead of the ratio of ca. 0.93 measured in the bulk \cite{Johansson1936}. This ratio, when applied to the bulk structure would lead to an interplane distance of 0.056~nm between the Cu and Au planes. In the actual 2D structure the distance between these planes is hence nearly four times larger to compensate for the much shorter $c$ as compared to bulk. Based on electronic band structure calculations (see Supplementary Information Figure S5), the 2D structure has a similar metallic character as bulk metals, and is not magnetic.
Finite temperature molecular dynamics simulations (1~fs time step, 300~K, 50 Cu and 50 Au atoms) reveal that the 2D CuAu structure is stable against thermal fluctuations.
However, there exists a buckled configuration that is energetically close to the flat ground state configuration (see Supplementary Information Figure S6). Thus, upon structural disturbances, 2D CuAu can have a tendency for buckling out of plane. This instability is observed experimentally during electron irradiation, as described below.
To further assess the stability of the 2D structure, we also simulated structures where some of the atoms of small (consisting of 70 or 100 atoms) islands were lifted from their in-plane positions on top of the structure. These structures had consistently higher energies than the perfect 2D structure, suggesting that the formation of a 2D island is preferred. We point out that a more comprehensive computational study, such as those in Refs. \cite{Cuddy2014, Nie2017} would demand a significant additional computational effort.

\subsection*{Phase transformation} Electron beam-driven structural changes are common during transmission electron microscopy, especially in thin structures, such as 2D materials~\cite{Susi2019}. Energy deposition from the energetic electrons can occur both indirectly due to electronic excitations or directly via elastic collisions between electrons and the atomic nuclei, of which only the latter is expected for good electronic conductors. At the acceleration voltages used here (60~kV and 80~kV), electrons can directly transfer a maximum of $\leq 1.0$~eV of kinetic energy to Au and $\leq 3.0$~eV to Cu atoms. In small gold clusters imaged at higher acceleration voltages (200 kV) electron irradiation has been shown to lead to continuous phase changes \cite{Smith1986}. Also the 2D CuAu structures are prone to electron beam damage despite the low transferred energies, but in a unique way (a typical experiment is shown in the Supplementary Information Figure S3 and Movie). This is likely due to the tendency for buckling that can make the structure locally unstable upon electron impacts. Indeed, already the lowest electron doses required for atomic resolution imaging tend to lead to changes at the edges of the structures where a clear reduction of area is visible while thicker regions start appearing. At the same time, small vacancy-type defects appear inside the structure, growing to larger defects with increasing electron irradiation dose.

\begin{figure*}[h]
	\includegraphics[width=\textwidth]{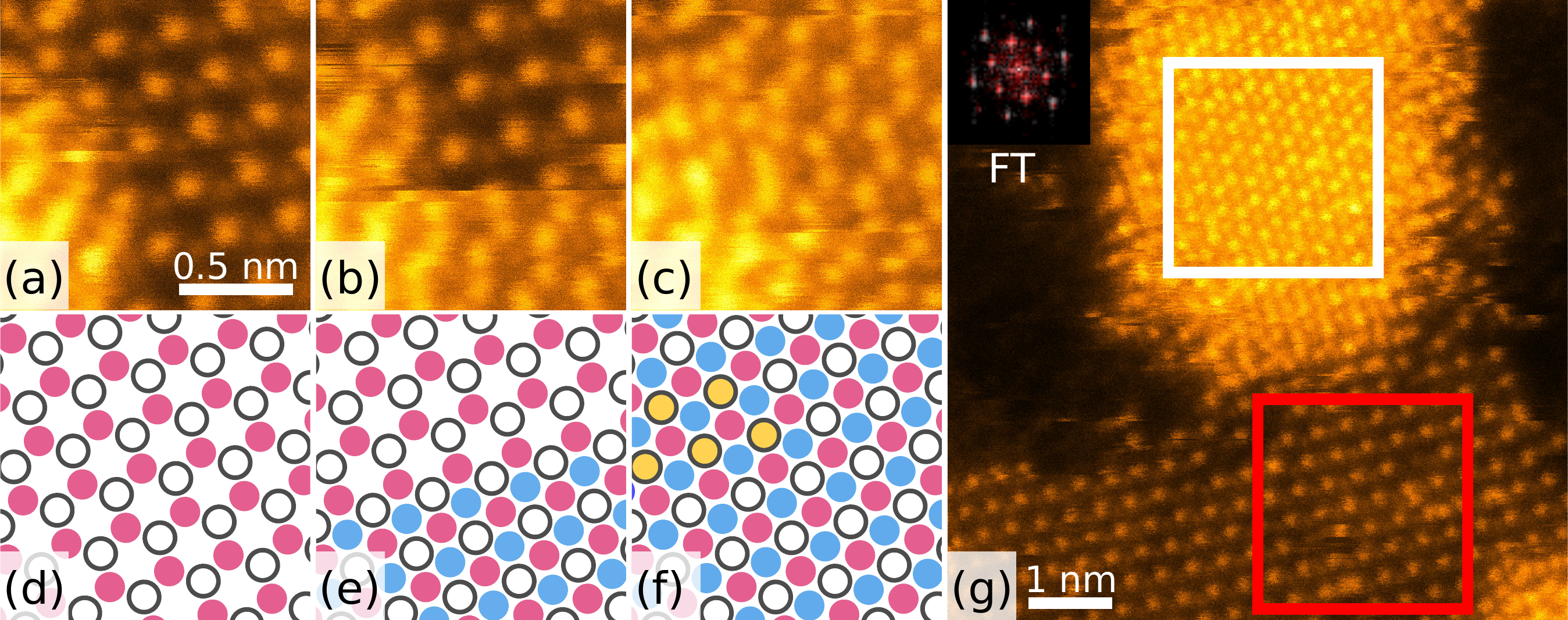}
	\caption{Phase transformation due to electron irradiation during observation. a) 2D CuAu next to a thicker region on the left.
		During the raster acquisition of the next image (b) the structure becomes partially covered with an additional layer of gold atoms, based on the HAADF intensity. c) The complete area is covered by the additional layer, and another gold layer has started to form on the left part of the image. d)-f) Schematic illustration of the atomic positions seen in the above images. Copper (hollow circle) and gold atoms (red circle) form the base of the hexagonal crystal. The first adlayer of gold atoms (blue circle) in (e) fills the holes in the hexagonal crystal. f) The second adlayer of gold atoms (golden circle) occupies the position above the copper atoms (empty circle) of the original structure. g) A cluster with four additional layers shows the same $\alpha$-$\beta$-$\gamma$ stacking, known from fcc crystals in the [111] direction. Fourier transforms (FT) of 2D CuAu (red) and the thicker structure (white) from the marked areas show differing lattice constants. Images were recorded at 60~kV.
	}
	\label{fig:Crumpling} 
\end{figure*}

To understand how the phase transformation proceeds, we recorded atomically resolved images (Figure \ref{fig:Crumpling}a-c) of the process for a monolayer region adjacent to an already formed thicker cluster. During the STEM image acquisition (16.5 seconds per image), a third layer of atoms is appearing on top of the initially two-atom-thick structure (Figure \ref{fig:Crumpling}b). The additional layer completes within the acquisition of the image, and a few atoms have already started to create a fourth atomic layer on top of the structure in Figure \ref{fig:Crumpling}c. The layer-by-layer growth follows a close packed structure, which for a one-component material would correspond to $\alpha-\beta-\gamma$ stacking in the [111] lattice direction of the fcc structure, as illustrated in Figure \ref{fig:Crumpling}d-f.
In agreement with the experiment, DFT calculations show that gold adlayers on CuAu are favourable over small three-dimensional gold clusters (see Supplementary Information Figure S7). However, as is clear from other experimental images (see Supplementary Information Figure S3), eventually the continued electron irradiation leads to more three-dimensional shapes and finally to the formation of nanoclusters.
In Figure \ref{fig:Crumpling}g the phase transformation of that part of the 2D island has finished after four additional layers have accumulated on top of the original structure. Comparing the Fourier transform (inset) of the cluster (white) to the monolayer region (red) shows that the lattice constant increases from 0.268~nm by 6\% to 0.284~nm due to the phase transformation. Based on the intensity of the atoms participating in this process, additional layers consist of gold. This can be understood based on the bulk CuAu structure, since its formation would require distorting the underlaying hexagonal structure. Bulk gold also has the fcc structure consistent with stacking during the phase transformation, with a projected lattice constant of 0.289~nm in the [111] direction. DFT calculations similarly found an increase in the lattice constant with increasing number of Au adlayers (to 0.274~nm for CuAu with two added gold layers). This process occurred for all imaged 2D structures with sizes above some tens of atoms, and serves hence as an additional way of detecting the structures quickly among three-dimensional nanoclusters in the microscope. In large scale images or at lower resolution (as when recording the EDX map, see Figure \ref{fig:EDX}d-g) comparing two subsequent images with each other often reveals either partially or completely transformed structures.

\section*{Conclusions}

We have demonstrated the growth of a two-dimensional material with metallic bonding. The material was grown via thermal evaporation of gold on free-standing graphene and hBN samples containing left-over copper from their own growth process, creating thus van der Waals heterostructures with each of the substrate materials. It consists of separate layers of copper and gold atoms, which each have a hexagonal intralayer symmetry, separated by 0.224~nm to form a honeycomb structure with an in-plane lattice constant of 0.268(2)~nm. The lattice is similar to two adjacent (201) atomic planes of the bulk $L1_0$ CuAu crystal, with the difference that for the honeycomb structure a $c/a$ ratio of exactly $1/\sqrt{2}$ is required instead of the bulk value of ca. 0.93 and the distance between the copper and gold layers is significantly larger. In the two-dimensional structure the spins are paired and no magnetism is predicted.
Upon irradiation with an electron beam, the two-dimensional material undergoes a phase transformation and forms three-dimensional clusters. Due to its metallic nature and atomic thinness it could be a beneficial electrode material for applications utilizing 2D materials and their heterostructures. Because the 2D structures anchor at contamination sites on the substrate, such structures could also be patterned via electron-beam-induced deposition \cite{Zagler2019}. Catalytic and electro-optical performances of Au and CuAu make the exploration of such properties for this resource-conserving and straightforwardly produced novel material intriguing.

\section*{Methods}

\subsection*{Sample Preparation}
Commercially available graphene (Graphenea Inc.) and hBN monolayers (Graphene Laboratories Inc.), grown by chemical vapour deposition on copper substrates, deposited on perforated SiN transmission electron microscopy grids were used as substrates for the growth. Graphene substrates were cleaned in the microscope column with a high-power laser (6~W) to obtain large clean areas. Connected in the same vacuum system, Au was deposited (Knudsen Cell thermal evaporator at 1080$^\circ$C) onto the substrates, with abundant Cu left on the substrate.

\subsection*{Microscopy and Spectroscopy}
Scanning transmission electron microscopy high-angle annular dark field images were acquired with a Nion UltraSTEM 100 in Vienna (annular range 80--300~mrad, convergence semiangle 30~mrad). Electron energy loss spectroscopy measurements were carried out in the same microscope with a Gatan PEELS 666 spectrometer, retrofitted with an Andor iXon 897 electron-multiplying charge-coupled device camera.
Energy-dispersive X-ray spectroscopy measurements were carried out at an image-side corrected FEI Titan 80-300 transmission electron microscope equipped with an EDAX spectrometer.

\subsection*{Simulations}
DFT simulations were performed using the Vienna Ab initio Software Package (VASP) \cite{Kresse1996,Kresse1996a}, adopting the strongly constrained and appropriately normed (SCAN) functional \cite{Sun2015,Sun2016} with an energy cut-off of 500 eV. Structural optimization was carried out for the unit cell containing two atoms.
Finite temperature molecular dynamics simulations and comparison of 2D and 3D adstructures were carried out for a 2D CuAu structure with 50 Cu and 50 Au atoms (with additional 12 Au atoms in the latter case).
STEM image simulations were carried out using PyQSTEM~\cite{Koch2002,Madsen2017}.

\section*{Acknowledgments} The authors thank A. Mittelberger and J. C. Meyer for their early contribution in testing gold evaporation onto graphene, which served as an inspiration for the study, and R. Singh for his help in the EDX measurements.

\section*{Funding} J.K. acknowledges financial support through the Austrian Science Fund via project I3181-N36.

\section*{Competing interests} The authors declare no competing interests.

\section*{Data availability} All data needed to evaluate the conclusions in the paper are present in the paper or the Supplementary Information. All original microscopy images and spectra are available through the data repository of the University of Vienna~\cite{phaidra}.

\section*{Supplementary Information} Supplementary Information contains electron energy loss spectra of Cu and CuAu, a complete atomic-resolution STEM-HAADF time series of the phase transformation of a 30~nm large 2D CuAu structure, EELS measurements for pre-characterization of a Cu cluster on graphene and for a 2D CuAu structure, an illustration of two adjacent atomic (120) planes of the CuAu $L1_0$ structure with $c/a = 1/\sqrt{2}$ that have the same hexagonal structure as the observed 2D CuAu, energy comparison for flat and buckled 2D CuAu structures, electronic band-structure calculations and an energy comparison between a small gold cluster and a gold adlayer on CuAu. Also, two videos illustrating the phase transformation and the rotation of 2D CuAu on a graphene substrate, confirming that it is not epitaxially grown on the substrate, are provided.

\clearpage

\section*{Supplementary Information}
\renewcommand{\figurename}{Supplementary Figure}
\renewcommand{\tablename}{Supplementary Table}
\setcounter{figure}{0}
\setcounter{page}{1}

\begin{figure*}[!htp]
	\includegraphics[width=\textwidth]{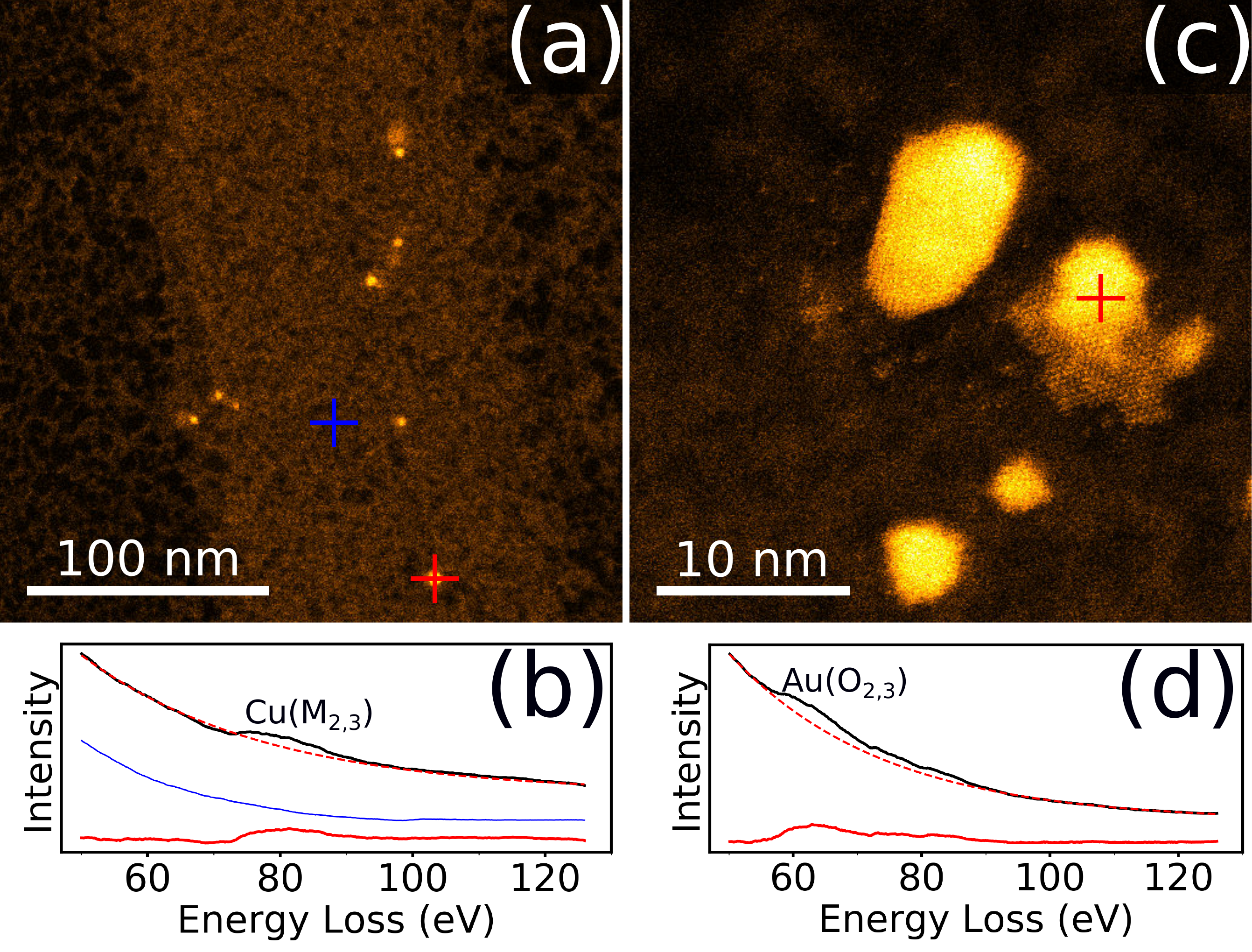}
	\caption{Electron energy loss spectroscopy before and after Au deposition. a) Graphene substrate before laser cleaning and Au deposition, covered in contamination with some brighter 3D clusters. b) The signal in black from the cluster marked by a red cross in (a) displays after background subtraction (red dashed line, yielding red solid line) a copper $M_{2,3}$ edge and no other peaks. The blue signal, taken at a region without a cluster (blue cross), shows no signal, indicating that the spectrometer is calibrated correctly. c) A cluster is formed from the phase transformation of 2D CuAu. d) EELS signal corresponding to the structure marked by red cross in (c) shows the gold $O_{2,3}$ edge. The Au $O_{2,3}$ peak together with $N_{6,7}$ at 83 eV extend over the weaker Cu $M_{2,3}$ making unambiguous identification of Cu in the alloy impossible. The black lines are raw spectra, dashed red lines the fitted background and red solid lines the resulting data after background subtraction. All data was recorded at 60~kV.
	}
	\label{fig:Cu_before_Au_dep} 
\end{figure*}

\begin{figure*}[!htp]
	\includegraphics[width=\textwidth]{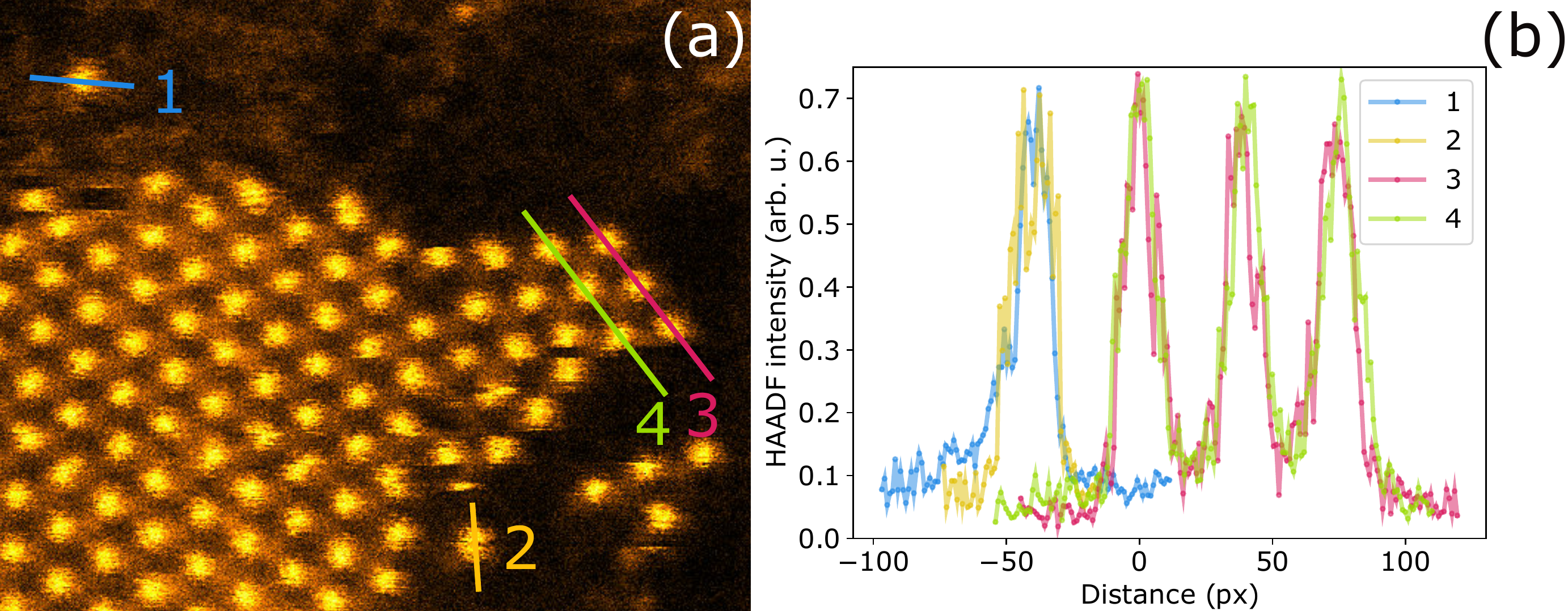}
	\caption{Comparison of the STEM HAADF signal of single atoms to atomic positions in hexagonal CuAu.
		a) Line profiles of the HAADF signal are taken of single atoms (1 and 2) and along atomic positions in CuAu (3 and 4).
		b) The peak of the HAADF intensity of single atoms (1 and 2) is comparable in height to the signal from the atomic positions in the 2D structure (3 and 4).
	}
	\label{fig:HAADF_comparison} 
\end{figure*}

\begin{figure*}[!htp]
	\includegraphics[width=\textwidth]{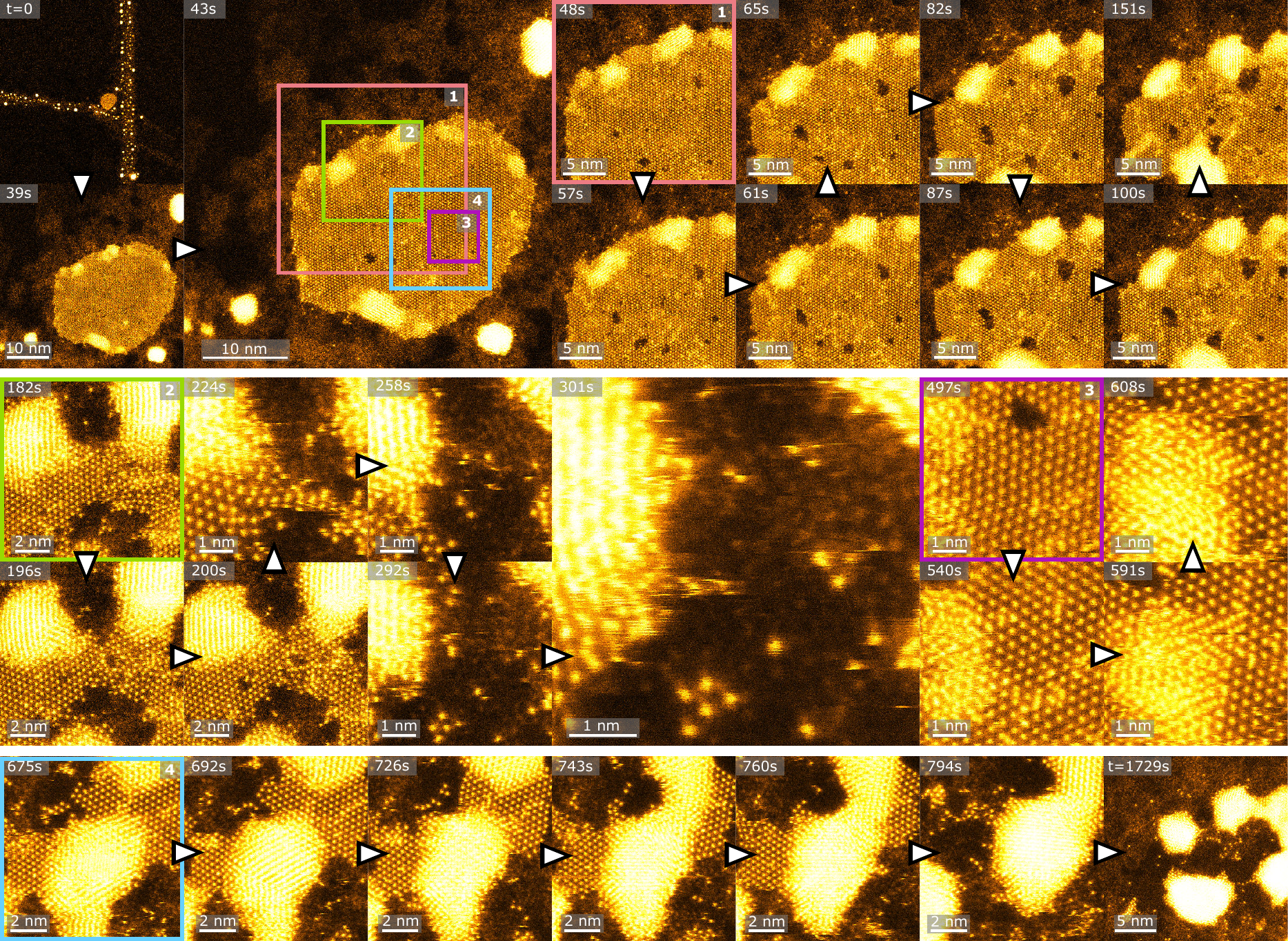}
	\caption{Electron irradiation effects on monolayer CuAu. Electron irradiation causes the material to undergo a phase transformation ($0-39$ s)---regions on the rim vanish from the interaction with the electron beam while at the same time brighter clusters appear on the 2D metal. Further irradiation ($48-100$ s) creates holes in the 2D material and enlarges them, while the clusters grow. In severely damaged 2D regions this process can be highly dynamic ($182-301$ s \& $726$ s) with sections disappearing during and between scans. Moving to another location on the 2D structure ($497$ s), reveals a mostly intact crystal. Defects in the crystal can also be healed ($497-540$ s), most likely from mobile Au and Cu from the process. Ultimately, the whole 2D structure transforms ($1729$ s) into a set of 3D nanoclusters. All images were recorded at 80~kV, with an approximate beam current of 30 pA.
	}
	\label{fig:Crumpling_long} 
\end{figure*}

\begin{figure*}[!htp]
	\includegraphics[width=\textwidth]{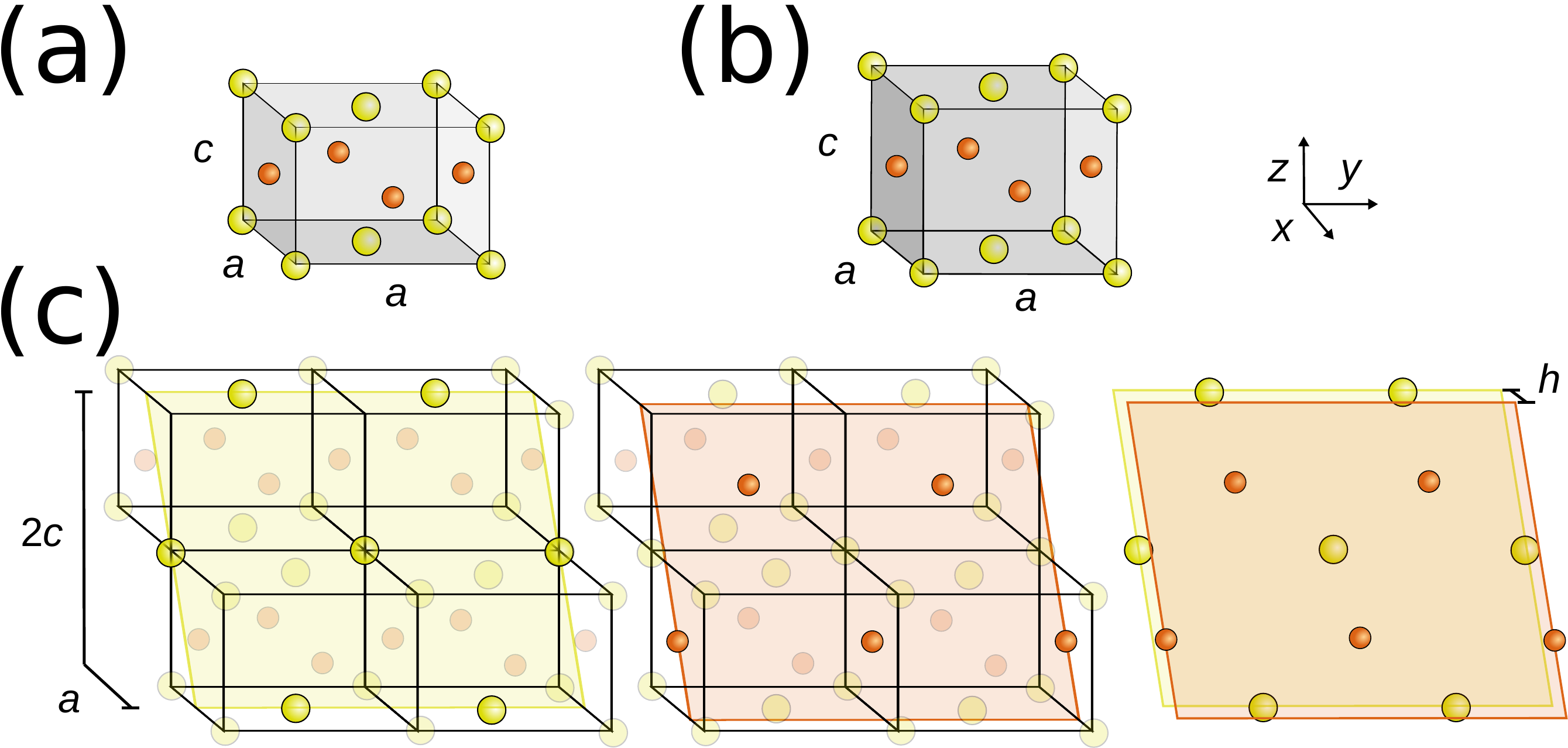}
	\caption{Structure of CuAu. a) Structure of tetragonal $L1_0$ CuAu with $c/a = 1/\sqrt{2}$ needed for a hexagonal crystal structure within the (201) plane. b) Structure of tetragonal $L1_0$ CuAu with $c/a = 0.93$ as obtained for the bulk structure experimentally (see main text).
		c) Two adjacent (201) planes for the structure of (a) corresponding to trigonal arrangements of Au in the first plane (yellow atoms) and Cu in the second one (orange atoms), and the atoms from those two planes shown together from the same perspective. The height difference between the Au and Cu planes as calculated from the structure is $h \approx 0.056$~nm, which is nearly four times shorter than what is obtained computationally for the actual 2D structure (0.224~nm).
	}
	\label{fig:CuAu_bulk_structure} 
\end{figure*}

\begin{figure*}[!htp]
	\includegraphics[width=\textwidth]{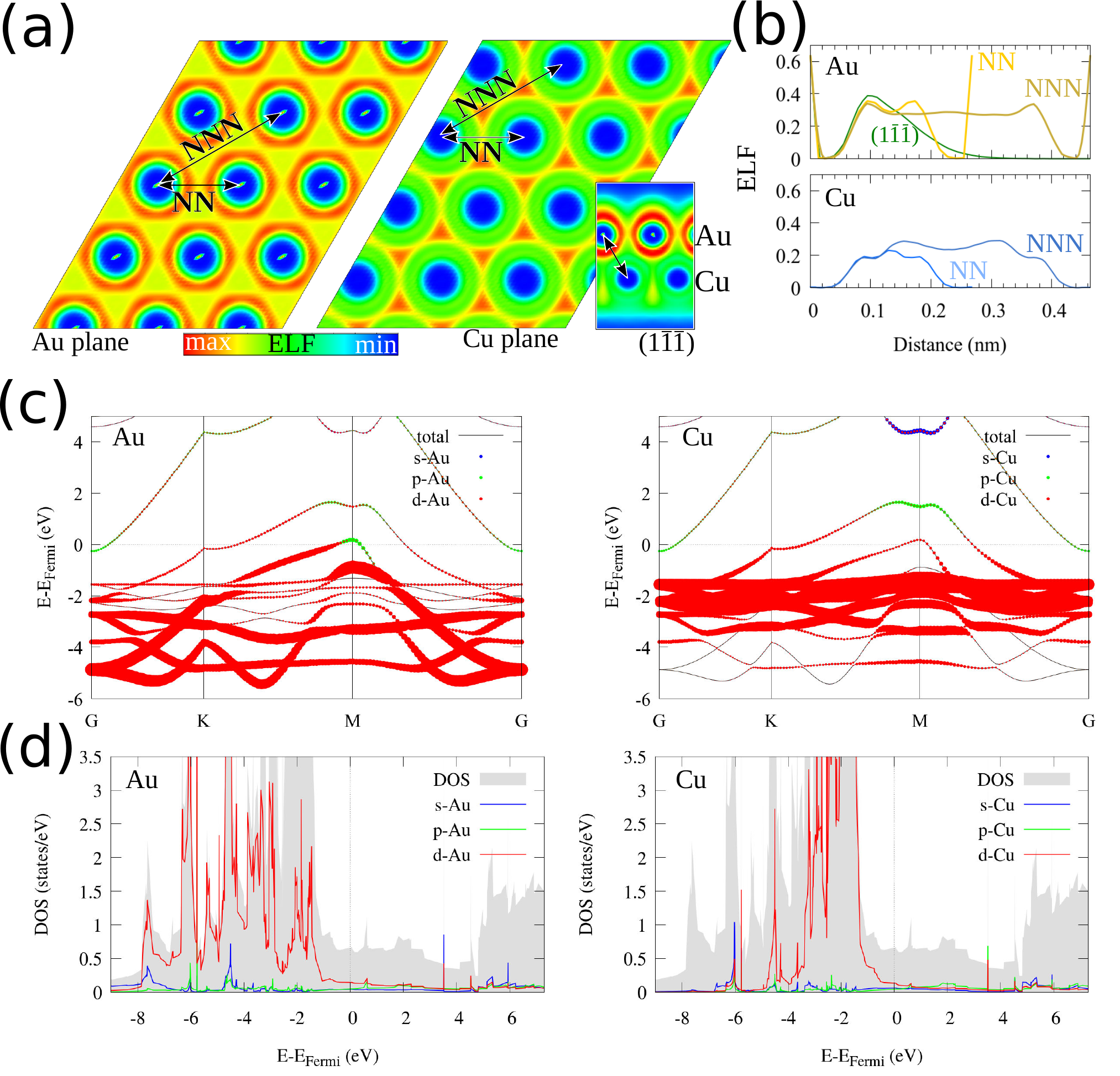}
	\caption{Electronic properties of hexagonal CuAu.
		a) Electronic localization function (ELF) as a contour plot on the Au (left), Cu (right) and the $(1\bar{1}\bar{1})$ fcc (inset) planes.
		b) ELF line profile between two nearest neighbor (NN) and two next-nearest neighbor (NNN) Au (top) or Cu (bottom) atoms, and between Au and Cu atoms on the $(1\bar{1}\bar{1})$ plane (top), as indicated by the arrows in (a).
		c) Fat band plots of the electronic band structures for Au and Cu states separately.
		d) Density of states, projected on the Au and Cu states. As evident from the figure, CuAu is metallic, with a majority of $d$-states crossing the Fermi level. The electronic bands show bad-metal -like characteristics, with shallow electron and hole pockets at G and M points, respectively. ELF shows a tendency for the electrons to assume a quite uniform distribution in the interstitial regions, as typical in metals.
	}
	\label{fig:electronics} 
\end{figure*}

\begin{figure*}[!htp]
	\includegraphics[width=\textwidth]{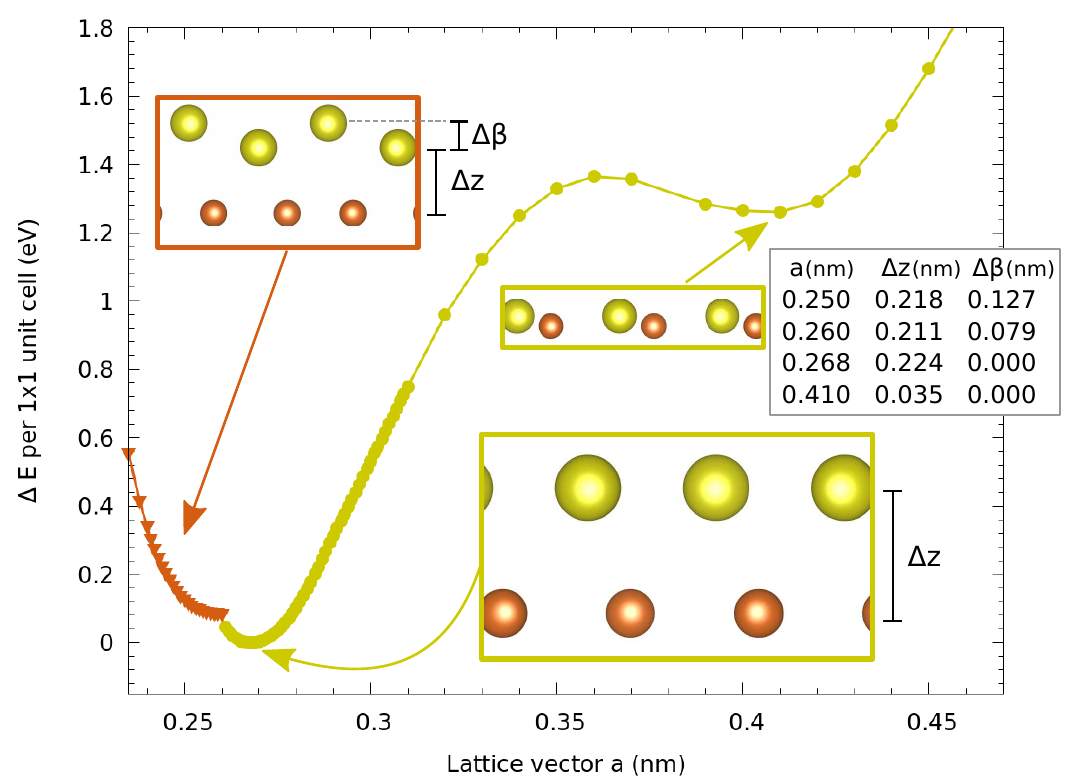}
	\caption{Buckling of the Au layer.
		DFT calculations performed on a super cell containing two Cu and two Au atoms show a competition between the flat and a buckled structure.
		The flat structure is more stable with a lattice constant of $a=0.268$ nm and an interlayer distance of $\Delta z= 0.224$ nm. By applying compressive strain, the structure is driven towards a phase transition: at $a=0.26$ nm, the Au atoms buckle out of the plane ($\Delta \beta = 0.079$ nm); larger compressive strains further enhance the buckling (see the table included in the figure). Conversely, the Cu layer remains flat upon compression. We relate this different behaviour to the different lattice constants of the Au and Cu free-standing monolayers (0.269 nm and 0.244 nm, respectively, as calculated by DFT). On the one hand, the small energy difference of $\Delta E= 80 $ meV between the buckled structure and the ground-state determine a dynamical instability at finite temperatures, as discussed in the main text.
		On the other hand, the flat structure is stable upon tensile strain, and it continuously decreases the interlayer distance up to stabilizing a quite flat monolayer of both Cu and Au atoms ($\Delta E = 1.25$ eV).
	}
	\label{fig:buckling} 
\end{figure*}

\begin{figure*}[!htp]
	\includegraphics[width=\textwidth]{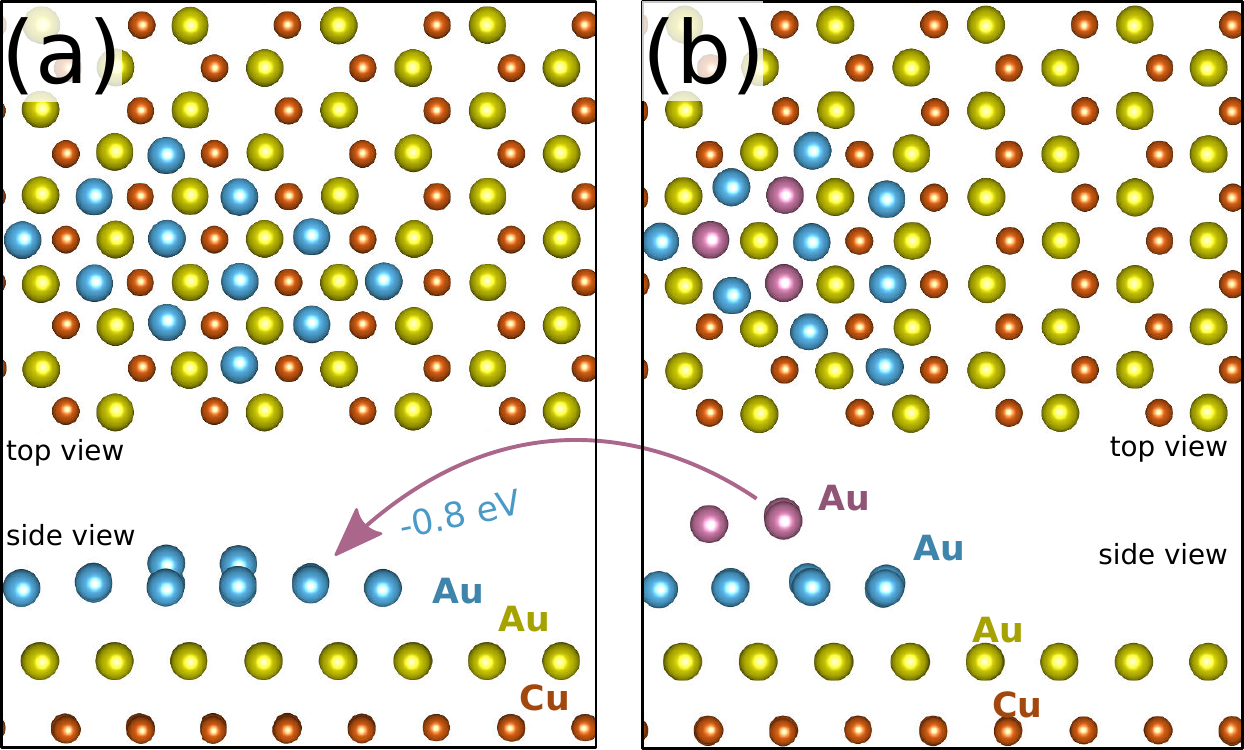}
	\caption{Tendency to form an additional layer instead of a cluster.
		DFT calculations using a 2D CuAu structure with 50 Cu and 50 Au atoms with additional 12 Au atoms arranged on top show that an arrangement corresponding to (a) one additional fcc (111)-layer is favoured over (b) a small cluster by $0.8$ eV.
	}
	\label{fig:clustering} 
\end{figure*}

\end{document}